\documentclass[preprint,showpacs,preprintnumbers,amsmath,amssymb]{revtex4}
\usepackage{graphicx,epsfig,dcolumn,bm,epic,eepic,float}
\usepackage{amsmath}
\usepackage{latexsym}
\usepackage{booktabs}
\usepackage{makeidx,shortvrb,latexsym}
\usepackage{color}
\begin{document}
\title{Application of the KMR and MRW unintegrated parton distributions to the EMC ratio of $^6{Li}$ nucleus   in the $k_t$-factorization  framework}
\author{M. Modarres }\altaffiliation {Corresponding author, Email:
mmodares@ut.ac.ir, Tel: +98-21-61118645, Fax: +98-21-88004781}
\author{A. Hadian }\altaffiliation {ahmad.hadian@ut.ac.ir}
\affiliation{Physics Department, University of  Tehran, 1439955961,
Tehran, Iran.}
\begin{abstract}
In the present work, it is intended to calculate the unintegrated
parton distribution functions (UPDFs) of the $^6{Li}$ nucleus, which
depend not only on the longitudinal momentum fraction $x$  (the
Bjorken variable) and the factorization scale $\mu$ of partons, but
also on their   transverse momentum ($k_t$). Therefore, the KMR and
MRW procedures  are applied to generate the $k_t$-dependent parton
distributions from the familiar integrated parton densities (PDFs),
which   were determined in our recent related work,  using the
constituent quark exchange model (CQEM) for the $^6Li$ nucleus.
Then, the resulting UPDFs of $^6{Li}$ are  compared with the UPDFs
of free proton from our previous work. Afterwards, from the
$k_t$-factorization formalism, the structure function (SF) of
$^6{Li}$ nucleus is computed  to extract its European Muon
Collaboration (EMC) ratio. The   results are compered with those
generated from the $^6Li$ nucleus PDFs and the available NMC
experimental data. It is observed that, especially in the small $x$
region, where the $k_t$ dependence of partons are important in the
calculations of the hadron structure functions, the present EMC
ratios are extremely improved compared with those of our previous
work, and show an excellent agreement with the NMC experimental
data.
\end{abstract}
\pacs{ 13.60.Hb, 21.45.+v, 14.20.Dh, 24.85.+P, 12.39.Ki\\ Keywords:
Unintegrated parton distribution function,  KMR and MRW frameworks,
Constituent quark exchange model, Structure function, EMC ratio.}
\maketitle
\section{Introduction}
Conventionally, deep inelastic scattering (DIS) processes are
analyzed to determine the parton distribution functions (PDFs) of
the different targets. These distributions correspond to the density
of partons in the parent hadron with longitudinal momentum fraction
$x$ (the Bjorken variable), integrated over the parton transverse
momentum up to $k_t$ = $\mu$. Thus, the usual PDFs are not
$k_t$-dependent distributions and they satisfy the standard
Dokshitzer-Gribov-Lipatov-Altarelli-Parisi (DGLAP) evolution
equations in the factorization scale $\mu$
\cite{Gribov,Lipatov,Altarelli1,Dokshitzer}. However, in the recent
years, it is found that especially at small $x$ region, the
transverse momentum of partons become important in the
electron-proton and proton-proton collisions. An enormous amount of
experiments are conducted at the high-energy particle physics
laboratories (such as the exclusive and semi-inclusive processes in
the high energy collisions in the LHC), which show that the parton
distributions unintegrated over $k_t$ are more appropriate. These
$k_t$-dependent parton distributions are called the unintegrated
parton distribution functions (UPDFs) and depend on two hard scales,
i.e., the factorization scale $\mu$ and the transverse momentum
$k_t$, which satisfy the much more complicated
Ciafaloni-Catani-Fiorani-Marchesini (CCFM) equations
\cite{Ciafaloni,Catani,Catani2,Marchesini,Marchesini2}.

The procedure of solving the CCFM equations is mathematically
intricate and impractically time consuming, since it includes
iterative integral equations with many terms. On the other hand, the
CCFM formalism can be exclusively used for the gluon evolution and
is impotent to produce a convincing quark contribution.  To overcome
these complications, a different approach based on the $LO$ DGLAP
evolution have been proposed by Kimber, Martin and Ryskin (KMR)
\cite{Kimber2}. Afterwards, to improve the exclusive processes,
Martin, Ryskin and Watt (MRW) extended the KMR formalism to the
leading order $(LO)$ and next-to-leading order $(NLO)$ levels
\cite{Martin1}.  These two procedures, i.e., the KMR and MRW, are
constructed by imposing the different angular ordering constraints
on the standard DGLAP equations and can produce UPDFs by using the
conventional integrated single-scale PDFs as inputs. Recently, we
  intensely applied the KMR and MRW prescriptions in the various
studies; see references
\cite{Modarres2,Modarres3,Modarres4,Modarres5,Modarres6,Modarres7,Modarres8,Modarres9,Modarres10,
Modarres11,Olanj1,Hosseinkhani1}. In the section III, we briefly
explain the concepts of KMR and MRW prescriptions.

As we pointed out before, it was shown that especially in the very
small $x$ region, the UPDFs play significant roles in the structure
functions (SFs) of free nucleons and nuclei
\cite{Kimber1,Kimber2,Olanj1,Hosseinkhani1,Oliveira1,Martin1}. It is
well known  that the   SFs of   free and bound nucleons in the
nuclear medium are not the same. In other words, the ratio of the
nuclear SF to that of the free nucleon deviates from unity. This
phenomenon was first declared in 1983 by the European Muon
Collaboration (EMC) group, i.e., Aubert et al \cite{Aubert1}, and is
referred to as the EMC effect (or EMC ratio). The utilization of
UPDFs to the nuclei was investigated by Martin group
\cite{Oliveira1}, which was demonstrated that the UPDFs can improve
the nucleus SF in the small $x$ domain. These outcomes motivate us
to modify our previous SFs and EMC ratios of $^6{Li}$ nucleus, by
considering the KMR and MRW UPDFs in our prior formalisms
\cite{Hadian1,Hadian2}.

As we mentioned, to produce the UPDFs, one  requires the usual
integrated PDFs, i.e., the distribution of quarks, $xq(x,Q^2)$, and
gluons, $xg(x,Q^2)$, as inputs. The single-scale PDFs of $^6{Li}$
nucleus at the hadronic scale $\mu_0^2$ = 0.34 $GeV^2$, can be
determined by applying the constituent quark exchange model (CQEM)
for the $A$ = 6 iso-scalar system, i.e. $^6Li$ nucleus,
\cite{Hadian1}, and
  they could be evolved to any required higher energy scale,
$\mu^2$,  by using  the DGLAP equations \cite{Botje}. Then,  these
PDFs  can be used as the inputs to the KMR  and MRW formalisms, to
extract the corresponding UPDFs. The summary of the CQEM for the
$^6Li$ nucleus will be presented in the section II.

So, the paper is organized as follows: First, a brief explanation of
the CQEM for the  $^6Li$ nucleus will be presented in the section II
and the appendix A. In the section III, the KMR and MRW approaches
for extraction of the double-scale UPDFs from the conventional bound
integrated PDFs will be reviewed. The formulation of the SF,
$F_2(x,Q^2)$, and the EMC ratio based on the $k_t$-factorization
formalism will be given in the section IV. Finally, the section V
will be devoted to the results, discussions, and conclusions.
\section{The CQEM for the $\mathcal{A}=6$ iso-scalar system}
Actually, the CQEM is constructed of two primary formalisms, i.e.,
the quark exchange framework (QEF) and the constituent quark model
(CQM). The QEF was primordially introduced by Hoodbhoy and Jaffe to
extract the valence quark distributions in the three nucleons mirror
nuclei \cite{Jaffe1,Hoodbhoy1}. Recently, this method was expanded
by us to calculate the constituent quark distribution functions of
$^6{Li}$ nucleus \cite{Hadian1,Hadian2}. Nevertheless, the QEF is
incapable of producing the sea quark and gluon distributions. So we
apply the CQM, which was first established by Feynman in 1972
\cite{Feynman,Close,Roberts}, to generate partons degrees of freedom
from the QEF  that gives only the valence quark distributions. We
simultaneously convolute  the QEF and the CQM to derive  the PDFs of
$^6{Li}$ nucleus and denote this combination as  the constituent
quark exchange model (CQEM i.e. CQM$\otimes$QEF) (see the reference
\cite{Hadian1}).
\subsection{The QEF for the $^6 Li$ nucleus}
First, according to the appendixes A, we tend to introduce the QEF
for the $^6 Li$ nucleus to produce the constituent quark
distributions of $^6{Li}$ nucleus. The quark momentum distribution
in the six-nucleon system can be written as follows:
\begin{equation}\label{1}
\rho(\vec{k},\mathcal{A}_i)={{\left< \mathcal{A}_i=6 \right|
q^\dagger_{\mu}q_{\mu} \left| \mathcal{A}_i=6 \right>}\over {\left<
\mathcal{A}_i=6 | \mathcal{A}_i=6 \right>}},
\end{equation}
where ${\left| \mathcal{A}_i=6 \right>}$ is the nucleus state and
$q_\mu^\dagger$ ($q_\mu$) indicates the creation (annihilation)
operator for the quarks with the state index $\mu$. After evaluating
relevant computations, which are given in appendix A in detail, the
final quark momentum density is obtained as
\begin{equation}\label{2}
\rho(k)={{\rho_{dir}(k)+\rho_{exch}(k)}\over
{\Big[1+{45\over8}\mathcal{I}\Big]}},
\end{equation}
where
$$\rho_{dir}(k)={9}A(k){\quad},$$
$$\rho_{exch}(k)={10}B(k)+{10\over3}C(k)+{5\over3}D(k).$$
The coefficients $\mathcal{I}, A, B, C,$ and $D$ are determined in
the equations (\ref{A.18}), and (\ref{A.20}) to (\ref{A.23}) of the
appendix A, respectively. One can clearly find out that, the
strength of the exchange term  is proportional to the coefficients
$B$, $C$, and $D$ and these coefficients  depend on the overlap
integral $\mathcal{I}$ which is a function of  the nucleon's radius.
The consistency of our quark density results for the six-nucleon
iso-scalar system can be easily verified via the following sum rule:
\begin{equation}\label{3}
\int{\rho(k)d{\vec{k}}}={18\over2}.
\end{equation}
The constituent quark distributions are determined from the quark
momentum distributions in the nucleon of the nucleus
$\mathcal{A}_i$, at each $Q^2$ scale via the following equation ($j$
= $p$, $n$ ($a$ = $u$, $d$) for the proton (up quark) and neutron
(down quark), respectively) \cite{Jaffe1}
\begin{equation}\label{4}
f^a_j(x,Q^2;\mathcal{A}_i)=\int{{{\rho^a_j(\vec{k};\mathcal{A}_i)}}\delta\Big({x-{k_+\over
M}}\Big)d{\vec{k}}},
\end{equation}
where the light-cone momentum of the constituent quark in the target
rest frame is used, and $k^0$ is considered as a function of
${|\vec{k}|}$
${({k^0=[{({{\vec{k}^2}+m^2}})}^{1\over2}-{\epsilon}_0})$. The
parameters ${m}$, and ${\epsilon}_0$ are defined as the quark mass
and its binding energy, respectively. we regard both as free
parameters to be fit to the valence up quark distribution of Martin
$et$ $al.$, i.e., MSTW 2008 \cite{Stirling,Stirling2,Stirling3}. So,
in the present work, their numerical values are chosen as $m$ = 320
MeV and $\epsilon_0$ = 120 MeV. After performing the angular
integration in the equation (\ref{4}), the constituent distribution
is finally obtained as follows,
\begin{equation}\label{5}
f^a_j(x,Q^2;\mathcal{A}_i)={{2\pi
M}}\int^{\infty}_{k^a_{min}}{{{\rho^a_j(\vec{k};\mathcal{A}_i)}}kdk},
\end{equation}
with,
\begin{equation}\label{6}
k^a_{min}(x)={({{x M}+{\epsilon^a_0})^2}-m^2_a\over{2({x
M}+{\epsilon^a_0})}},
\end{equation}
where $M$ denotes the nucleon mass.
\subsection{The CQM for the $^6Li$ nucleus}
Now, to complete the CQEM, we present a brief description to the
concept of CQM. In this approach, the main idea is that the
constituent quarks are themselves complex objects, whose structure
functions are determined by a set of functions $\phi_{ab}(x)$ that
define the number of point-like partons of type $b$ inside the
constituent of type $a$ with fraction $x$ of its total momentum.
These functions for various kinds of partons were defined in the
references
\cite{Altarelli,Scopetta1,Manohar,Vento,Yazdanpanah1,Yazdanpanah2}.
Therefore, the fundamental equation of this framework is specified
as follows:
\begin{equation}\label{7}
q(x,\mu_0^2)=\int^{1}_{x}{dz\over
z}\Big[{\mathcal{U}(z,\mu_0^2)\phi_{\mathcal{U}q}\Big({x\over z},
\mu_0^2\Big)+\mathcal{D}(z,\mu_0^2)\phi_{\mathcal{D}q}\Big({x\over
z}, \mu_0^2\Big)}\Big],
\end{equation}
where $q$ labels the various point-like partons, i.e., valence
quarks ($u_v$, $d_v$), sea quarks ($u_s$, $d_s$, $s$), sea
anti-quarks ($\bar{u}_s$, $\bar{d}_s$, $\bar{s}$) and gluons ($g$).
The $\mathcal{U}$ and $\mathcal{D}$ represent the constituent
density distributions of $u$ and $d$ quarks, respectively, and can
be obtained from QEF, i.e., the equation (\ref{5}). The $\mu_0^2$ =
0.34 $GeV^2$ is the hadronic scale at which the CQM is defined.
Based on the CQM, the sea quark and anti-quark distributions are
independent of iso-spin flavor. So in what follows, we demonstrate
these distributions with $q_s$. It should be noted that in the
constituent quark of type $\mathcal{U}$, there is no point-like
valence quark of type $d$ and vice versa. Therefore, the functions
$\phi_{\mathcal{U}d}\Big({x\over z},\mu_0^2\Big)$ and
$\phi_{\mathcal{D}u}\Big({x\over z}, \mu_0^2\Big)$ are omitted from
calculations. Additionally, for the $^6{Li}$ iso-scalar nucleus, the
constituent distributions of $u$ and $d$ quarks are equal, i.e.,
$\mathcal{U}(z,\mu_0^2)$ = $\mathcal{D}(z,\mu_0^2)$.

Based on the above primary introduction, the single-scale point-like
parton distributions of $^6Li$ nucleus at the hadronic scale
$\mu_0^2$ can be obtained from the CQEM according to the following
equations:
\begin{equation}\label{8}
u_v(x,\mu_0^2)=d_v(x,\mu_0^2)=\int^{1}_{x}{dz\over
z}{\mathcal{U}(z,\mu_0^2)\phi_{\mathcal{U}q_v}\Big({x\over z},
\mu_0^2\Big)},
\end{equation}
\begin{equation}\label{9}
q_s(x,\mu_0^2)=2\int^{1}_{x}{dz\over
z}{\mathcal{U}(z,\mu_0^2)\phi_{\mathcal{U}q_s}\Big({x\over z},
\mu_0^2\Big)},
\end{equation}
\begin{equation}\label{10}
g(x,\mu_0^2)=2\int^{1}_{x}{dz\over
z}{\mathcal{U}(z,\mu_0^2)\phi_{\mathcal{U}g}\Big({x\over z},
\mu_0^2\Big)},
\end{equation}
where
\begin{equation}\label{11}
\mathcal{U}(z,\mu_0^2)={{2\pi
M}}\int^{\infty}_{k_{min}}{{{\rho(k)}}kdk},
\end{equation}
with
$$\rho(k)={{{9}A(k)+{10}B(k)+{10\over3}C(k)+{5\over3}D(k)}\over
{\Big[1+{45\over8}\mathcal{I}\Big]}}.$$
The  resulted PDFs are shown
in the figure 1. We take the numerical value of nucleon's radius as
$b$ = 0.8 fm throughout of present calculations. Subsequently, the
overlap integral should take the corresponding numerical value as
$\mathcal{I}$ = 0.0504 \cite{Jaffe1}.

As it was mentioned before, using the CQEM, the PDFs can be produced
only at the hadronic energy scale, $\mu_0^2$ = 0.34 $GeV^2$.
However, it should be noted that, given the PDFs at the some
reference point, $a(x,Q_0^2)$, we can compute them for any value of
$Q^2$ using the DGLAP equations. Therefore, to generate these
resulted PDFs, shown in the figure 1, at the higher energy scale
$Q^2$, the DGLAP evolution equation will be applied
\cite{Gribov,Lipatov,Altarelli1,Dokshitzer},
\begin{equation}\label{12}
{d\:a(x,Q^2)\over {d\:log(Q^2)}}=
{\alpha_s(Q^2)\over{2\pi}}\sum_{b=q,g}\Big[\int^{1}_{x}dzP_{ab}(z)\:
b\Big({x\over z},Q^2\Big)-a(x,Q^2)\int^{1}_{0}dzz\:P_{ba}(z)\Big].
\end{equation}
The splitting functions, $P_{ab}(z)$, account for the probability of
a parton of type $a$ with momentum fraction $x^{\prime\prime}$,
$a(x^{\prime\prime},Q^2)$, emerging from a parent parton of type $b$
with a larger momentum fraction $x^\prime$, $b(x^\prime,Q^2)$,
through $z = x^{\prime\prime}/x^\prime$. However, the DGLAP
evolution equation is based on the strong ordering assumption, which
systematically neglects the transverse momentum of the emitted
partons along the evolution ladder. Therefore, in the following
section, the KMR and MRW methods, which are based on the DGLAP
equations along with some modifications, will be given to consider
the transverse momentum of the parton distributions explicitly.
\section{The KMR and MRW formalisms}
The KMR framework as well as the $LO$ and the $NLO$ MRW approaches
are briefly presented in the following two subsections.
\subsection{The KMR procedure}
 The KMR formalism was
introduced by Kimber, Martin and Ryskin \cite{Kimber2,Kimber1}. They
modified the DGLAP equations by separating the real and virtual
contributions of the evolution at the $LO$ level and defined the
two-scale UPDFs, $f_a(x,k_t^2,\mu^2)$, where $a$ = $q$ or $g$, as
follows:
\begin{equation}\label{13}
f_a(x,k_t^2,\mu^2) = T_a(k_t^2,\mu^2)\sum\limits_{b=q,g}\Big[
{\alpha_s(k_t^2)\over{2\pi}}\int^{1-\Delta}_{x}dzP_{ab}^{(LO)}(z)b\Big({x\over
z},k_t^2\Big)\Big],
\end{equation}
where $P_{ab}^{(LO)}$ represent the $LO$ splitting functions and the
survival probability $T_a$ is given by
\begin{equation}\label{14}
T_a(k_t^2,\mu^2)=exp\Big(-\int^{\mu^2}_{k_t^2}{\alpha_s(k^2)\over{2\pi}}{dk^2\over
k^2}\sum\limits_{b=q,g}\int^{1-\Delta}_{0}{dz^\prime}P_{ab}^{(LO)}(z^\prime)\Big),
\end{equation}
which gives the probability that parton $a$ with transverse momentum
$k_t$ remains untouched in the evolution up to the factorization
scale $\mu$. The infrared cut-off, $\Delta$ = $1-z_{max}$ =
$k_t/(\mu+k_t$), is introduced via imposing the angular ordering
condition (AOC) on the last step of the evolution, and protects the
$1/(1-z)$ singularity in the splitting functions arising from the
soft gluon emission. In the above formulation, the key observation
is that the dependence on the second scale $\mu$ of the UPDFs enters
only at the last step of the evolution.  The cut-off $\Delta$ in the
KMR formalism is imposed to both the quark and gluon terms. While
this cut-off is generated from AOC which theoretically perceivable
for terms including the gluon emissions, i.e., the diagonal
splitting functions $P_{qq}(z)$ and $P_{gg}(z)$.
\subsection{The MRW procedure}
The $LO$ MRW scheme was defined by Martin, Ryskin and Watt as a
correction to the KMR framework and shortly afterwards, was expanded
into the $NLO$ level \cite{Martin1}. In the rest of this section the
concepts of both the $LO$ and $NLO$ MRW approaches will be
presented.

The general forms of UPDFs of the $LO$ MRW for the quarks and gluons
are given in the equations (\ref{15}) and (\ref{17}), respectively:
\begin{align}\label{15}
f_q^{LO}(x,k_t^2,\mu^2)=T_q(k_t^2,\mu^2){\alpha_s(k_t^2)\over{2\pi}}\int^{1}_{x}dz\Big[P_{qq}^{(LO)}(z){x\over
z}q\Big({x\over z},&k_t^2\Big)\Theta\Big({\mu\over \mu+k_t}-z\Big)
\nonumber\\&+\:\:P_{qg}^{(LO)}(z){x\over z}g\Big({x\over
z},k_t^2\Big)\Big],
\end{align}
where
\begin{equation}\label{16}
T_q(k_t^2,\mu^2)=exp\Big(-\int^{\mu^2}_{k_t^2}{\alpha_s(k^2)\over{2\pi}}{dk^2\over
k^2}\int^{z_{max}}_{0}{dz^\prime}P_{qq}^{(LO)}(z^\prime)\Big),
\end{equation}
and
\begin{align}\label{17}
f_g^{LO}(x,k_t^2,\mu^2)=T_g(k_t^2,\mu^2){\alpha_s(k_t^2)\over{2\pi}}\int^{1}_{x}dz\Big[&\sum_{q}P_{gq}^{(LO)}(z){x\over
z}q\Big({x\over
z},k_t^2\Big)\nonumber\\&+\:\:P_{gg}^{(LO)}(z){x\over
z}g\Big({x\over z},k_t^2\Big)\Theta\Big({\mu\over
\mu+k_t}-z\Big)\Big],
\end{align}
where
\begin{equation}\label{18}
T_g(k_t^2,\mu^2)=exp\Big(-\int^{\mu^2}_{k_t^2}{\alpha_s(k^2)\over{2\pi}}{dk^2\over
k^2}\Big[\int^{z_{max}}_{z_{min}}{dz^\prime}z^\prime
P_{qq}^{(LO)}(z^\prime)+n_f\int_0^1{dz^\prime}P_{qg}^{(LO)}(z^\prime)\Big]\Big).
\end{equation}
The upper limit of the integration on the variable $z$ is defined as
$z_{max}=1-z_{min}=\mu/(\mu+k_t)$, and $n_f$ is the flavor number.
In the present study, we consider three lightest flavor of quarks,
i.e., $u$, $d$ and $s$. So, $n_f$ = 3 throughout of our
calculations.

The $LO$ UPDFs of MRW formalism can be expanded into the $NLO$
region according to the following equations:
\begin{equation}\label{19}
f_a^{NLO}(x,k_t^2,\mu^2)=\int_x^1dzT_a(k^2,\mu^2){\alpha_s(k^2)\over{2\pi}}\sum_{b=q,g}\tilde{P}_{ab}^{(LO+NLO)}(z)\:\:
b^{NLO}\Big({x\over
z},k^2\Big)\Theta\Big(1-z-{k_t^2\over\mu^2}\Big),
\end{equation}
where
\begin{equation}\label{19-b}
\tilde{P}^{(LO+NLO)}=\tilde{P}^{(LO)}+{(\alpha_s/{2\pi})}\tilde{P}^{(NLO)},
\end{equation}
and
\begin{equation}\label{20}
\tilde{P}_{ab}^{(i)}(z)=P_{ab}^{(i)}(z)-\Theta(z-(1-\Delta))\:\delta_{ab}F_{ab}^iP_{ab}(z).
\end{equation}
Here $i = 0, 1$ denote the $LO$ and $NLO$ contributions,
respectively. More details about the $NLO$ splitting functions are
given in the references \cite{Martin1,Furmanski}. It should be noted
that in the equation (\ref{19}), the parton transverse momentum
$k_t$ is related to the virtuality scale $k^2$ via the following
equation:
\begin{equation}\label{21}
k^2={k_t^2\over (1-z)}.
\end{equation}
In addition, the correct AOC for the soft gluon emission is provided
via the theta function $\Theta(z-(1-\Delta))$, and $\Delta$ can be
defined as
\begin{equation}\label{22}
\Delta={\sqrt{k^2(1-z)}\over {\mu+\sqrt{k^2(1-z)}}}\:\:.
\end{equation}
The final point is to present the Sudakov form factors $T_a$ at the
$NLO$ level, which again resume the virtual DGLAP contributions
during the evolution from $k^2$ to $\mu^2$, via the following
equations:
\begin{equation}\label{23}
T_q(k^2,\mu^2)=exp\:\Big(-\int^{\mu^2}_{k^2}{\alpha_s(\kappa^2)\over{2\pi}}{d\kappa^2\over
\kappa^2}\int^{1}_{0}{dz^\prime}z^\prime[\tilde{P}_{qq}^{(LO+NLO)}(z^\prime)+\tilde{P}_{gq}^{(LO+NLO)}(z^\prime)]\Big),
\end{equation}
\begin{equation}\label{24}
T_g(k^2,\mu^2)=exp\:\Big(-\int^{\mu^2}_{k^2}{\alpha_s(\kappa^2)\over{2\pi}}{d\kappa^2\over
\kappa^2}\int^{1}_{0}{dz^\prime}z^\prime[\tilde{P}_{gg}^{(LO+NLO)}(z^\prime)+2n_f\tilde{P}_{qg}^{(LO+NLO)}(z^\prime)]\Big).
\end{equation}
It was shown that by regarding only the $LO$ part of the complete
splitting functions, which were defined in the equation
(\ref{19-b}), the reasonable $NLO$ UPDFs with considerable accuracy
would be achieved \cite{Martin1}.

Now, by completing the procedures of generating the UPDFs from each
of above methods, i.e, the KMR, $LO$ MRW, and $NLO$ MRW, we can
compute the UPDFs of the $^6Li$ nucleus by using the conventional
single-scale bound PDFs, which previously were determined in section
III, as the inputs. These resulted UPDFs, $f_a(x,k_t^2,\mu^2)$, can
be interpreted as the probability of finding a parton of type $a$,
which carries the fraction $x$ of longitudinal momentum of its
parent hadron and with the transverse momentum $k_t$, in the scale
$\mu$ at the semihard level of a particular deep inelastic
scattering process. In the following section, we will present the
formulation of the deep inelastic SF, $F_2(x,Q^2)$, in the
$k_t$-factorization framework for the $^6Li$ nucleus.
\section{The SF and the EMC ratio calculation from the UPDFs in the $k_t$ factorization framework}
To check the reliability of our UPDFs, we briefly describe how to
use these distributions in calculations of the SF, $F_2(x,Q^2)$
\cite{Kimber2,Kimber1}. We explicitly investigate the separate
contributions of gluons and (direct) quarks to SF expression.

The gluons can only contribute to $F_2$ via an intermediate quark.
There are both the quark box and crossed-box diagrams of the figure
2, which must be regarded as the unintegrated gluon contributions.
The variable $z$ is used to denote the fraction of the gluon's
momentum that is transferred to the exchanged struck quark. As shown
in the figure 2, the parameters $k_t$ and $\kappa_t$ define the
transverse momentum of the parent gluons and daughter quarks,
respectively. The unintegrated gluon contributions to $F_2$ in the
$k_t$-factorization framework can be written as follows
\cite{Kimber1,Kimber2,Kwiecinski,Askew,Stasto}:
\begin{align}\label{25}
F_2^{g\rightarrow q\bar{q}}(x&,Q^2)=\sum_q\:e_q^2{Q^2\over
4\pi}\int{dk_t^2\over {k_t^4}}\int_0^1 d\beta\int
d^2\kappa_t\:\alpha_s(\mu^2)\:f_g\Big({x\over
z},k_t^2,\mu^2\Big)\Theta\Big(1-{x\over
z}\Big)\nonumber\\&\times\Big\lbrace[\beta^2+{(1-\beta)}^2]\:\Big({\boldsymbol{\kappa}_t
\over D_1}-{{\boldsymbol{\kappa}_t-\boldsymbol{k}_t}\over
D_2}\Big)^2\:+\:[m_q^2+4Q^2\beta^2(1-\beta)^2]\Big({1\over
D_1}-{1\over D_2}\Big)^2\Big\rbrace.
\end{align}
The variable $\beta$ is defined as the light-cone fraction of the
photon's momentum carried by the internal quark line and the
denominator factors are
\begin{align}\label{26}
&D_1=\kappa_t^2+\beta(1-\beta)Q^2+m_q^2\nonumber\\&D_2=(\boldsymbol{\kappa}_t-\boldsymbol{k}_t)^2+\beta(1-\beta)Q^2+m_q^2.
\end{align}
The summation is over various quark flavors $q$ which can appear in
the box, with different masses $m_q$. As we mentioned before, in
this work we consider the three lightest quark flavors ($u$, $d$,
$s$), which with a good approximation, their masses are neglected.
The variable $z$, which is the ratio of Bjorken variable $x$ and the
fraction of the proton momentum carried by the gluon, is specified
as
\begin{equation}\label{27}
{1\over
z}=1\:+\:{{\kappa_t^2+m_q^2}\over{(1-\beta)Q^2}}\:+\:{{k_t^2+\kappa_t^2-2\:\boldsymbol{\kappa}_t\:.\:\boldsymbol{k}_t+m_q^2}\over
{\beta Q^2}}.
\end{equation}
Following the reference \cite{Kwiecinski}, the scale $\mu$, which
controls the unintegrated gluon distribution and the QCD coupling
constant $\alpha_s$, is chosen as follows:
\begin{equation}\label{28}
\mu^2=k_t^2+\kappa_t^2+m_q^2.
\end{equation}
The equation (\ref{25}) gives the contributions of the unintegrated
gluons to $F_2$ in the perturbative region, $k_t$ $>$ $k_0$, where
the UPDFs are defined. The smallest cutoff $k_0$ can be chosen as
the initial scale of order 1 $GeV$, at which the $k_t$-factorization
theorem derives \cite{Askew}. The contributions of nonpertubative
region for the gluons, $k_t$ $<$ $k_0$, can be approximated such
that:
\begin{align}\label{29}
\int_0^{k_0^2}{dk_t^2 \over
k_t^2}\:f_g(x,k_t^2,\mu^2)\:\Big[{\textit{remainder\:\:of\:\:equation}\:\:(\ref{25})\over
k_t^2}\Big]\simeq\:xg(x,k_0^2)\:T_g(k_0,\mu)\:\Big[\quad\Big]_{k_t=a},
\end{align}
where $a$ is taken to be any value in the interval (0, $k_0$), which
its value is numerically unimportant to the nonperturbative
contributions.

Now we aim to add the contributions of unintegrated quarks to $F_2$.
Suppose that an initial quark with Bjorken scale $x$/$z$ and
perturbative transverse momentum $k_t$ $>$ $k_0$, splits into a
radiated gluon and a quark with smaller Bjorken scale $x$ and
transverse momentum $\kappa_t$. This final quark can interact with
the photon and contributes to $F_2$, as follows:
\begin{align}\label{30}
F_2^{q(perturbative)}(x,Q^2)=\sum_{q=u,d,s} \:e_q^2
\:\int_{k_0^2}^{Q^2}\:&{d\kappa_t^2\over
\kappa_t^2}\:{\alpha_s(\kappa_t^2)\over{2\pi}}\int_{k_0^2}^{\kappa_t^2}\:{dk_t^2\over
k_t^2}\int_x^{Q/{(Q+k_t)}}\:dz\nonumber\\&\times\Big[f_q
\Big({x\over z},k_t^2,Q^2 \Big)\:+\:f_{\bar{q}}\: \Big({x\over
z},k_t^2,Q^2 \Big)\Big]P_{qq}(z),
\end{align}
where the AOC during the quark evolution is imposed on the upper
limit of the $z$ integration.

Again, the nonperturbative contributions must be accounted for the
domain $k_t$ $<$ $k_0$,
\begin{equation}\label{31}
F_2^{q(nonperturbative)}(x,Q^2)=\sum_q
e_q^2\:\Big(xq(x,k_0^2)\:+\:x\bar{q}(x,k_0^2)\Big)\:T_q(k_0,Q),
\end{equation}
which physically can be interpreted as a quark (or antiquark) that
does not experience real splitting in the perturbative region, but
interacts unchanged with the photon at the scale $Q$. Therefore, a
Sudakov-like factor, $T_q(k_0, Q)$, is written to represent the
probability of evolution from $k_0$ to $Q$ without any radiation.

Eventually, the total SF can be obtained by the sum of both gluon
and quark contributions. The resulted formula will be applied by us
to calculate the SF of $^6Li$ nucleus in the $k_t$-factorization
framework. In addition, the EMC ratio, which is the ratio of the SF
of the bound nucleon to that of the free nucleon, will be calculated
via the following equation:
\begin{equation}\label{32}
\mathcal{R}_{EMC}={F_2^T(x)\over{F^{T^\star}_2(x)}},
\end{equation}
where $T$ is the target averaged over nuclear spin and iso-spin and
$T^\star$ is a hypothetical target with exactly the same quantum
numbers but in which the nucleons are forbidden to make any quark
exchange \cite{Jaffe1}. So, by setting the overlap integral
$\mathcal{I}$ equal to zero in the equation (\ref{2}), the momentum
distribution of free nucleon would be produced. It should be noted
that, the effects of nuclear Fermi motion are excluded from both $T$
and $T^\star$. We use the $k_t$-factorization approach to calculate
the UPDFs, SF and subsequently, EMC ratio of $^6Li$ nucleus and
remarkable outcomes are obtained, which will be presented in the
next section.
\section{Results, discussions, and conclusions}
After such a brief introduction to the KMR and LO and NLO MRW
formalisms, we now tend to start the numerical UPDFs calculation of
the $^6Li$ nucleus. Then, the SF and EMC ratio of $^6Li$ nucleus are
computed using these resulted UPDFs.

The gluon and the up quark double-scale UPDFs of $^6Li$ nucleus at
scales $\mu^2$ = 27 and 100 $GeV^2$ are plotted in the left and
right panels of the figure 3, respectively. The double-scale UPDFs
are obtained using the KMR (the full curves), the LO MRW (the dotted
curves), and the NLO MRW (the dash curves) schemes. These UPDFs are
plotted at the transverse momentums $k_t^2$ = 0.4$\mu^2$ and
0.9$\mu^2$. The values of $\mu^2$ and $k_t^2$ are chosen such that
the present outcomes to be comparable with results were obtained in
the reference \cite{Olanj1} for the free proton, in which the MSTW
2008-NLO set of PDFs \cite{Stirling} were used as the inputs (these
comparisons will be presented in the figures 4 and 5). For better
comparison of the KMR and MRW prescriptions, the same NLO PDFs are
 used as the inputs. Comparing the left and right panels of
the figure 3, it can be seen that at a fixed transverse momentum,
i.e., $k_t^2$ = 0.4$\mu^2$ or 0.9$\mu^2$, by increasing the
factorization scale $\mu^2$ from 27 to 100 $GeV^2$ in each row, the
output UPDFs do not change considerably. On the other hand, by
increasing the transverse momentum $k_t^2$ from 0.4$\mu^2$ to
0.9$\mu^2$ along each column, unlike the KMR (the full curves) and
the LO MRW (the dotted curves) cases, there are a sizable decrease
in the NLO MRW (the dash curves) UPDFs. This effect is more
prominent in the case of up quarks. Therefore, the reduction of NLO
UPDFs are more sensitive to the variation of the $k_t$, than the
scale $\mu^2$. Additionally, as the Bjorken scale $x$ in each
diagram increases, the discrepancies between UPDFs, which resulted
from various methods, are suppressed. Therefore, the growth of $k_t$
and reduction of $x$, which are characteristics of the high energy
and $k_t$-factorization region, affect the output UPDFs
significantly. The same conclusions were made in the reference
\cite{Olanj1} for the gluon and up quark UPDFs of the free proton.
It should be noted that, although the same NLO integrated PDFs are
used in both the LO and NLO MRW prescriptions as inputs, but the
results obtained from these frameworks are completely different
(compare the dotted and the dash curves). As it was mentioned in the
reference \cite{Martin1} for the free proton, these discrepancies
arise because in the NLO level, we impose the appropriate scale,
namely $k^2$ = $k_t^2$/($1$ $-$ $z$). However in the LO prescription
we do not care about the precise scale and scales $k_t^2$ and $k^2$
are both acceptable.

The comparisons of the gluon and up quark UPDFs of $^6Li$ nucleus in
the KMR and NLO MRW approaches (previously shown in the figure 3)
with those of the free proton (KMR-MSTW and NLO MRW-MSTW) which were
obtained in the reference \cite{Olanj1}, are displayed in the
figures 4 and 5, respectively. The values of factorization scale
$\mu^2$ and the transverse momentum $k_t^2$ are the same as those
mentioned in the previous paragraph. It can bee seen that the gluon
and up quark UPDFs of $^6Li$ nucleus generally behave similar to
those of the free proton. In the figure 4, both the gluons KMR and
NLO MRW UPDFs of $^6Li$ nucleus are located below those of free
proton. These discrepancies are relatively sizable in the small $x$
region, which show not only the significant role of the gluons at
the low $x$ domain, but also the importance     of the
$k_t$-factorization contribution at the small $x$ region. However,
these differences reduce when the variable $x$ increases.

In the figure 5, by considering the first row diagrams, it can be
observed that the up quark UPDFs of $^6Li$ nucleus in both the KMR
and NLO MRW schemes at the scales $\mu^2$ = 27 $GeV^2$ and $k_t^2$ =
0.4$\mu^2$, are obviously different from those of the free proton.
However these differences become smaller due to increase of the
factorization scale $\mu^2$ to 100 $GeV^2$ or intensifying   the
transverse momentum $k_t^2$ to 0.9$\mu^2$ (compare discrepancies
between up UPDFs shown in the diagrams of the first row, with those
shown in the diagrams of the second and third rows). Therefore, at
the scales $\mu^2$ = 100 $GeV^2$ and $k_t^2$ = 0.9$\mu^2$, the
resulted up quark UPDFs of $^6Li$ nucleus in both of the KMR and NLO
MRW approaches, become very similar to those of the free proton.

The resulting SFs of the $^6Li$ nucleus in the $k_t$-factorization
framework, using the KMR and  LO and  NLO MRW UPDFs to be inserted
in the $F_2$ equations, i.e. (\ref{25}) and (\ref{30}), at energy
scales $Q^2$ = 4.5, 15 and 27 $GeV^2$, are plotted in the figures 6,
7 and 8, respectively. In the first row of each figure, the
"gluon-originated" contributions are shown as the dash curves and
the "quark-originated" parts are shown as the dotted curves. In
addition, the continuous curves, which are the sum of the gluon and
quark contributions, represent the total SFs of the $^6Li$ nucleus.
To make the results more comparable, in the second row of each
figure, the overall SFs values in the KMR (the full curve), the LO
MRW (the dotted curve), and the NLO MRW (the dash curve) approaches
are plotted again. While the behavior of the LO MRW curves are very
similar to the KMR results, the NLO MRW outcomes demonstrate
different behavior from both the KMR and the LO MRW cases. The same
conclusion  was made about  the longitudinal SF ($F_L$) of the free
proton in the reference \cite{Hosseinkhani1}. As shown in the figure
6, the main contribution to the $F_2$ at the energy scale $Q^2$ =
4.5 $GeV^2$, comes from the quark contributions. However, by
increasing the energy scale $Q^2$ to 15 and 27 $GeV^2$, the gluon
contributions increase, and  at the $Q^2$ =27 $GeV^2$, the gluon
contributions at the small $x$ region become more important than the
quark portions. Although,  one can see that the "gluon-drive" curves
fall steeply as one goes to the larger $x$ domain. So,  at larger
$x$ values, again, the main portion of $F_2$ comes  from the quark
contributions. As expected, by increasing the $Q^2$, the
recognizable rise in $F_2$ at the smaller values of $x$ occurs.

The comparison of the $^6Li$  SFs (the full curve) in the KMR
prescriptions at the energy scales $Q^2$ = 4.5, 15 and 27 $GeV^2$
(previously shown in the figures 6, 7 and 8 respectively) with those
of the free proton (KMR-MSTW 2008) using the MSTW 2008 PDFs as
inputs (the dash curves),  are exhibited in the panels (a), (b) and
(c) of the figure 9, respectively. The total SFs of a hypothetical
$^6Li$ target without any quark exchange between its nucleons (by
setting the exchange integral $\mathcal{I}$ equal to zero in the
momentum density formula, equation (2)), i.e., hypothetical free
nucleon, in the KMR prescription are also plotted in this figure for
comparison (the dotted curves). We consider the three lightest
flavors of quarks, i.e., $u$, $d$ and $s$, to calculate the SFs of
both the $^6Li$ nucleus and the free proton. One observes that the
SFs of our hypothetical free nucleon are in good agreement with the
SFs of the free proton (see the dash and the dotted curves in each
panel), especially at low $x$ region, where approximately there is
no effect of the valence quarks. Additionally, according to the
equation (33), the EMC ratio in the KMR approach at each energy
scale, can be obtained by considering the ratio of the full curve
($^6Li$ SF) to the dotted curve (hypothetical free nucleon SF). This
ratio is plotted in each panel of the figure 10 via the full curve.

The EMC ratios of $^6Li$ nucleus by considering the
$k_t$-factorization method at the energy scales 4.5, 15 and 27
$GeV^2$,  are plotted in the panels (a), (b) and (c) of the figure
9, respectively. The input UPDFs are provided via the KMR (the full
curves), the LO MRW (the dotted curves), and the NLO MRW (the dash
curves) schemes. Due to neglecting the Fermi motion, the EMC ratios
monotonically decline and the growth in the EMC ratios at the large
values of $x$ do not occur. Therefore, the EMC ratios are plotted in
$x$ $\leq$ 0.8 domain. In the panel (a), the dotted-dash curve
illustrates the $^6Li$ EMC ratio at $Q^2$ = 0.34 $GeV^2$ and $b$ =
0.8 $fm$, that presented from our prior work \cite{Hadian1}, in
which we ignored the UPDF contributions in the EMC computations. The
filled circles in each panel indicate the NMC experimental data of
the EMC ratios of $^6Li$ nucleus measured in deep inelastic
muon-nucleus scattering at a nominal incident muon energy of 200
$GeV$ \cite{Malace,Arneodo}. It is obvious that by employing the
$k_t$-factorization theory, the present EMC outcomes at the small
$x$ region are outstandingly improved with respect to our previous
work \cite{Hadian1}. While the $^6Li$ SFs obtained from the KMR and
MRW procedures at each $Q^2$   are completely recognizable (see the
full, dotted and dash curves in the second rows of the figures 6, 7
and 8), the $^6Li$ EMC ratios resulted from these prescriptions are
approximately the same, and as it is seen in the each panel of the
figure 10, the solid, dotted and dash curves overlap. The main point
is that, in the EMC calculations, the ratio of bound and free
nucleon SFs is considered. So, by regarding the $k_t$-factorization
property in each of the KMR, LO or NLO MRW approaches, this ratio
remains almost unchanged. By increasing the variable $x$, the
differences between our present and prior EMC ratios decrease, that
show the prominent contributions of UPDFs in SFs and EMC
calculations at the small $x$ region.   It should be noted that in
this work, our main aim is to concentrate on the experimental data
in the small $x$ range, i.e., 0.00014 $\leq$ $x$ $\leq$ 0.0125,
which was omitted in our previous works \cite{Hadian1,Hadian2} and
it is usually referred to as "shadowing effect" \cite{67}. The
corresponding NMC energy scales for that $x$ area span a very wide
range of low $Q^2$ (0.034 $\le$ $Q^2$ $\le$ 1.8 $GeV^2$). Therefore,
these small $x$ NMC data locate in the nonperturbative region and we
cannot apply the perturbative UPDFs in the EMC calculations for
them, individually. On the other hand, it is well known that the EMC
ratios are not $Q^2$ dependent, significantly (e.g. see reference
\cite{68}). As it can be easily seen in the figure 10, this point is
also established in our EMC calculations, and the resulting EMC
ratios at the energy scales 4.5, 15, and 27 $GeV^2$ are not very
different. So, by choosing these mentioned $Q^2$ values, we can
consider the perturbative contributions of UPDFs in the EMC
calculations. On the other hand,   the present outcomes in which the
$k_T$ dependence of partons takes into account, with respect to
 our previous work \cite{Hadian2},  reproduces the general form of the shadowing effect
\cite{67} at the small $x$ values, which was previously absent
\cite{Hadian1}.

In conclusion, we   employed the KMR, LO MRW, and NLO MRW frameworks
to elicit the two-scale unintegrated parton distribution functions
of $^6Li$ nucleus from the single-scale PDFs, which were generated
from the constituent quark exchange model at the hadronic scale 0.34
$GeV^2$ and  were evolved by the DGLAP evolution equations to
required higher energy scales. Subsequently, the resulted UPDFs were
compared with those of the free proton of our previous work
\cite{Olanj1} at the typical factorization scales $Q^2$ = 27 $GeV^2$
and 100 $GeV^2$, and desirable conclusions were presented.
Afterwards, the structure functions of $^6Li$ nucleus in the
$k_t$-factorization formalism were computed at the energy scales
$Q^2$ = 4.5, 15 and 27 $GeV^2$ using the UPDFs of KMR and LO and NLO
MRW prescriptions. Again, we have compared the SFs of $^6Li$ nucleus
with those of free proton. Eventually, the EMC ratios of $^6Li$
nucleus in the $k_t$-factorization scheme were calculated, and
compared with the NMC experimental data \cite{Malace,Arneodo} as
well as our corresponding previous work \cite{Hadian1}, in which we
neglected the contributions of UPDFs in EMC computations.

It should be noted that although the LO and the NLO MRW approaches
are more compatible with the DGLAP evolution equation and they were
introduced as extensions and improvements to the KMR formalism, but
based on our previous works (see for example references
\cite{Modarres7,Modarres8,Olanj1,Hosseinkhani1}, the KMR procedure
have better agreement with the experimental data. This is of course
due to the use of the different implementation of the AOC in the KMR
approach, which automatically includes the re-summation of
$ln(1/x)$, Balitski-Fadin-Kuraev-Lipatov (BFKL)
\cite{62,63,64,65,66} logarithms, in the LO DGLAP evolution
equation. In other words, the particular form of the AOC in the KMR
formalism, despite being of the LO level, includes some
contributions from the NLO sector, whereas in the MRW frameworks,
these contributions must be inserted separately. In other words,
because of these formulations,  the LO-MRW  and NLO-MRW formalisms
constraint quarks or gluons radiation in the LO and NLO levels,
respectively, while the KMR  approach constraints both quarks and
gluons radiations. So, compared to the MRW frameworks, the KMR
approach leads to the more precise results in the calculation of
different structure functions and cross sections, also see
\cite{AMIN,LIPIPPP} and the references therein. However in the
fragmentation regions, this conclusion may not be true \cite{AMIN1}.
So, Because of the above properties of the KMR approach, most of the
new works considered only the KMR approach. However, in the present
work, the only available experimental data is the EMC ratio of
$^6Li$ nucleus (not the SFs, itself) and, when one considers the
ratio of the bound to the free nucleon SFs theoretically, the SFs
errors will be canceled in the EMC division formula. Therefore, all
of the KMR, LO and NLO MRW approaches, with a good approximation,
give the same EMC results. However, we expect that if the
experimental data of the $^6Li$ SFs are reported in future, our
calculations in the KMR scheme in accordance to the above
conclusion, would be more consistent with the data. Finally, at the
very high momentum transverse and very small $x$, after the points
which were raised in the reference \cite{61}, these approaches need
further investigations in this region which cause the $k_t^2$
becomes greater than our hard scale. We hope we could make a final
comment about this point in our near future works.

As stated earlier, by considering the $k_t$-factorization approach,
the results were significantly improved in the small $x$ region and
the outcomes astonishingly were consistent with the NMC data.
Finally we should remark that     the inclusion of  $k_t$ dependent
PDFs in our EMC calculation, explains the  reduction of EMC effect
at the small $x$ region, which is traditionally known as  the
"shadowing phenomena" \cite{67,69,70}.  Some of  models presented in
the references \cite{67,69,70}, could be equivalent to our UPDFs
inclusion, e.g.  the Pomeron and Reggeon contributions to the
$\gamma^*p$ diffraction \cite{70} which can in general explain the
raise of PDFs at small the $x$ region \cite{71} in the framework of
the regge theory. However, we hope in our future works we could
consider these models in our results, as well.
\section*{Acknowledgements}
$MM$ would like to acknowledge the Research Council of University of
Tehran and Institute for Research and Planning in Higher Education
for the grants provided for him.
\begin{appendix}
\section{The QEF for the six-nucleon iso-scalar system}
Now, we describe the quark exchange model for $\mathcal{A}=6$
iso-scalar system in detail. It should be noted that, all
assumptions that have been made in the references
\cite{Owns,Hoodbhoy1,Jaffe1,Modarres,Chekanov} are valid here.
Especially, the Fermi motion is ignored by regarding the leading
order expansion of the nuclear wave function. In addition, since the
atomic number is still small ($\mathcal{A}$=6), as usual  we neglect
the possible simultaneous exchange of quarks between more than two
nucleons, which can be important as one moves to the heavy nuclei.
Additionally, because we do not have the full Lithium nucleus wave
function, and to make the calculations less complicated, the Lithium
nucleus is considered as a uniform nucleus system and all
calculations are performed at the nuclear matter density.  The
single nucleon  three valence quarks state is written as,
\cite{Betz,Jaffe1,Owns,Hoodbhoy1,Modarres,GHAFOORI},
\begin{equation}\tag{A.1}\label{A.1}
\left| \alpha
\right>={\mathcal{N}^{\alpha^\dagger}}\left|0\right>={1\over\sqrt{3!}}\mathcal{N}^\alpha_{\mu_1\mu_2\mu_3}q^\dagger_{\mu_1}
q^\dagger_{\mu_2}q^\dagger_{\mu_3}\left|0\right>,
\end{equation}
where the indices $\alpha$ $(\mu_i)$ describe the nucleon (quark)
states $\{\vec{p},M_S,M_T\}$ $(\{\vec{k},m_s,m_t,c\})$ (note that
$M_T$ $(m_t)$ = $+{1\over 2}$ and $-{1\over 2}$ for the proton
(up-quark) and neutron (down-quark), respectively). $q_\mu^\dagger$
$({\mathcal{N}^{\alpha^\dagger}})$ denote the creation operators for
the quarks (nucleons) with   a summation over repeated indices  as
well as integration over $\vec{k}$ is assumed. The totally
antisymmetric nucleon wave function is,
\begin{equation}\tag{A.2}\label{A.2}
\mathcal{N}^\alpha_{\mu_1\mu_2\mu_3} =
D(\mu_1,\mu_2,\mu_3;\alpha_i){\times}\delta(\vec{k_1}+\vec{k_2}+\vec{k_3}-\vec{P})\varphi(\vec{k_1},\vec{k_2},\vec{k_3},\vec{P}),
\end{equation}
where $\varphi(\vec{k_1},\vec{k_2},\vec{k_3},\vec{P})$, i.e. the
nucleon wave function, is approximated  as (b is the nucleon's
radius),
\begin{equation}\tag{A.3}\label{A.3}
{\varphi(\vec{k_1},\vec{k_2},\vec{k_3},\vec{P})}=\Big({3b^4\over\pi^2}\Big)^{3\over4}exp{\Big[{-b^2(k_1^2+k_2^2+k_3^2)\over2}}+{{b^2P^2}\over6}\Big],
\end{equation}
with ($C^{{j_1}{j_2}{j}}_{{m_1}{m_2}{m}}$ are Clebsch-Gordon
coefficients and $\epsilon_{{c_1}{c_2}{c_3}}$ is the color factor),
\begin{align}\label{A.4}
D(\mu_1,\mu_2,\mu_3;\alpha_i)=&{1\over\sqrt{3!}}\epsilon_{{c_1}{c_2}{c_3}}{1\over\sqrt{2}}\sum\limits_{s,t=0,1}
C^{{1\over2}{s}{1\over2}}_{{m_{s_\sigma}}{m_s}{M_{S_{\alpha_i}}}}\nonumber\\&{\quad}{\times}C^{{1\over2}{1\over2}{s}}_{{m_{s_\mu}}{{m_{s_\nu}}}{m_s}}
C^{{1\over2}{t}{1\over2}}_{{m_{t_\sigma}}{m_t}{M_{T_{\alpha_i}}}}C^{{1\over2}{1\over2}{t}}_{{m_{t_\mu}}{{m_{t_\nu}}}{m_t}}.
\tag{A.4}
\end{align}

The nucleus states are defined as,
\begin{equation}\tag{A.5}\label{A.5}
\left|\mathcal{A}_i=6 \right>
={1\over\sqrt{6!}}\chi^{{\alpha_1}{\alpha_2}{\alpha_3}{\alpha_4}{\alpha_5}{\alpha_6}}\mathcal{N}^{\alpha_1^\dagger}\mathcal{N}^{\alpha_2^\dagger}
\mathcal{N}^{\alpha_3^\dagger}\mathcal{N}^{\alpha_4^\dagger}\mathcal{N}^{\alpha_5^\dagger}\mathcal{N}^{\alpha_6^\dagger}\left|
0 \right>.
\end{equation}
$\chi^{{\alpha_1}{\alpha_2}{\alpha_3}{\alpha_4}{\alpha_5}{\alpha_6}}$
is the complete antisymmetric nuclear wave function
  where taken from the reference \cite{GHAFOORI}. Afnan et al.
\cite{Afnan,Bissey} found that the choice of nucleon-nucleon
potential or nuclear wave function does not dramatically   affect
the EMC results.

Then, the constituent quark momentum distributions   can be written
as follows:
\begin{equation}\tag{A.6}\label{A.6}
\rho_{m_t}^{M_T}(\vec{k},\mathcal{A}_i)={{\left< \mathcal{A}_i=6
\right| q^\dagger_{\mu}q_{\mu} \left| \mathcal{A}_i=6 \right>}\over
{\left< \mathcal{A}_i=6 | \mathcal{A}_i=6 \right>}}.
\end{equation}
Now by using the equation (\ref{A.6}) and performing some lengthy
algebra \cite{Hadian2}, one can find that the momentum distribution
in the $^6Li$ nucleus, $(\rho(k)$,  is   an iso-scalar distribution,
i.e., averaged on ${M}_{T}$) as follows:
\begin{equation}\tag{A.19}\label{A.19}
\rho(k)={{\rho_{dir}(k)+\rho_{exch}(k)}\over
{\Big[1+{45\over8}\mathcal{I}\Big]}},
\end{equation}
where
$$\rho_{dir}(k)={9}A(k),$$
$$\rho_{exch}(k)={10}B(k)+{10\over3}C(k)+{5\over3}D(k),$$
with
\begin{equation}\tag{A.20}\label{A.20}
A(k)=\Big({3b^2\over2\pi}\Big)^{3\over2}exp{\Big[-{3\over2}b^2{k^2}\Big]},
\end{equation}
\begin{equation}\tag{A.21}\label{A.21}
{\quad}B(k)=\Big({27b^2\over8\pi}\Big)^{3\over2}exp{\Big[-{3\over2}b^2{k^2}\Big]}\mathcal{I},
\end{equation}
\begin{equation}\tag{A.22}\label{A.22}
{\quad}{\quad}C(k)=\Big({27b^2\over7\pi}\Big)^{3\over2}exp{\Big[-{12\over7}b^2{k^2}\Big]}\mathcal{I},
\end{equation}
\begin{equation}\tag{A.23}\label{A.23}
{\quad}D(k)=\Big({27b^2\over4\pi}\Big)^{3\over2}exp{\Big[-{3}b^2{k^2}\Big]}\mathcal{I}.
\end{equation}
and
\begin{align}\label{A.18}
&\mathcal{I}=8\pi^2\int^{\infty}_{0}{x^2dx}\int^{\infty}_{0}{y^2dy}\nonumber\\&{\qquad}{\times}\int^{1}_{-1}{x^2dx}\int^{1}_{-1}d(cos{\theta})
{exp{\Big[-{{{3}x^2}\over{4b^2}}\Big]}}{\mid{\chi(x,y,cos{\theta})}\mid}^2.
\tag{A.18}
\end{align}
The same approximation as the one used in the references
\cite{Betz,Jaffe1,Owns,Hoodbhoy1,GHAFOORI} is applied, especially
the leading order expansion for $\chi(\vec{p},\vec{q})$ \cite{Betz}.
This approximation is equivalent to the omission of the Fermi motion
and will affect the structure function for $x\geq 0.75$. As we
pointed out before, according to references
\cite{Afnan,Bissey,Chen.,Stadler}, the other choices of
nucleon-nucleon potentials, do not considerably change the nuclear
wave functions and the EMC results.

 As we pointed out before, since we do
not have a complete lithium nucleus wave function and to reduce
further complications, the integral $\mathcal{I}$, which is defined
in the equation (\ref{A.18}), is calculated from the reference
\cite{GHAFOORI} at the nuclear matter density.  However, as shown in
our previous works, the results are not very sensitive to the choice
of $\mathcal{I}$ (its variation with respect to the different wave
functions is less than 1\% \cite{Jaffe1}).
\end{appendix}

\begin{figure}[h!]
  \includegraphics [ scale=0.6]{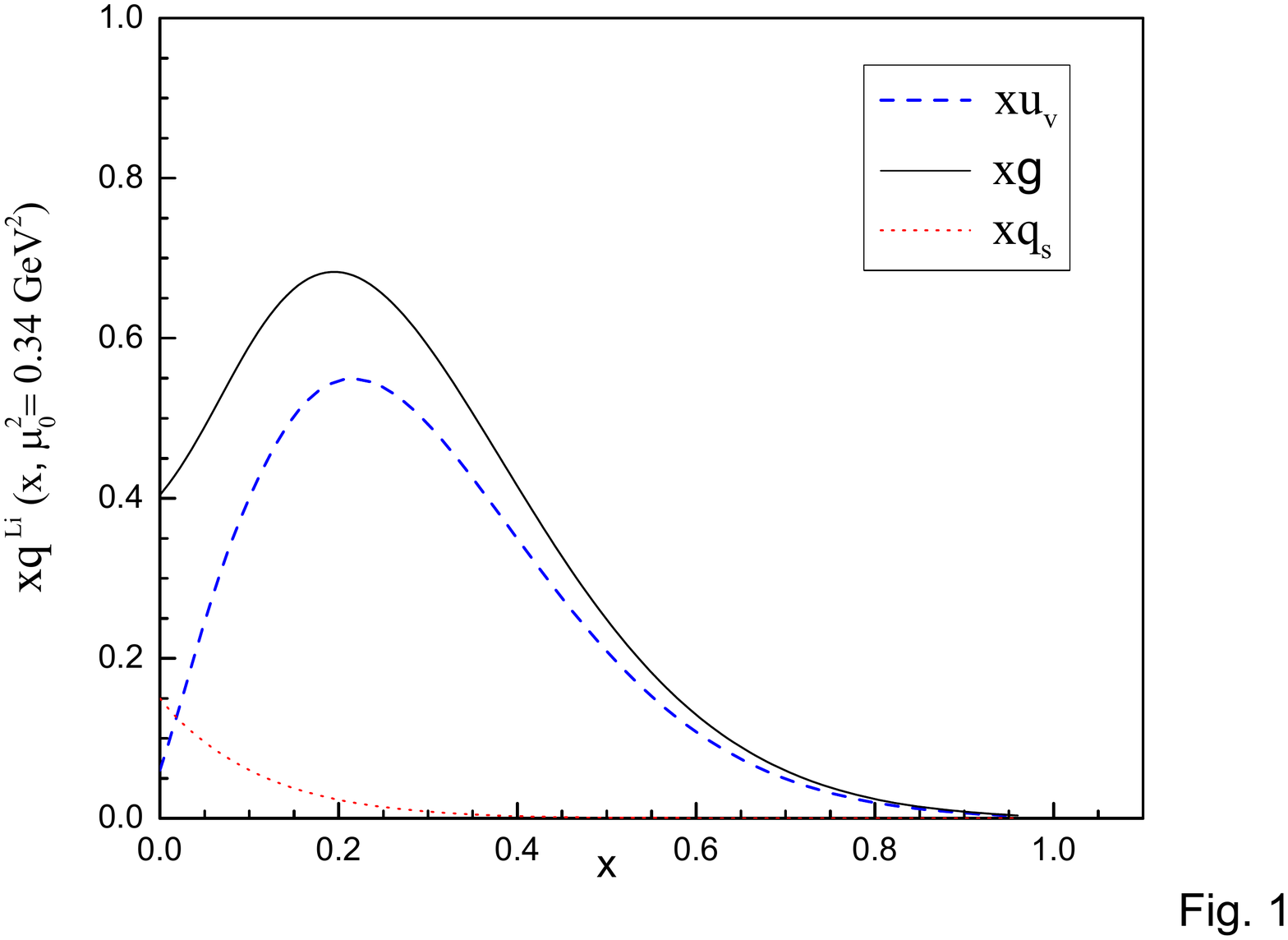}
\caption{The parton distribution functions of $^6Li$ nucleus versus
$x$, for $(m, \epsilon_0)$ pairs of (320, 120 $MeV$) and $b$ = 0.8
$fm$ at the hadronic scale, $\mu_0^2$ = 0.34 $GeV^2$. The full curve
represents the gluon distributions, while the dash and the dotted
curves indicate the valence and the sea quark distributions,
respectively.}
\end{figure}
\begin{figure}[h!]
  \includegraphics [ width=\linewidth]{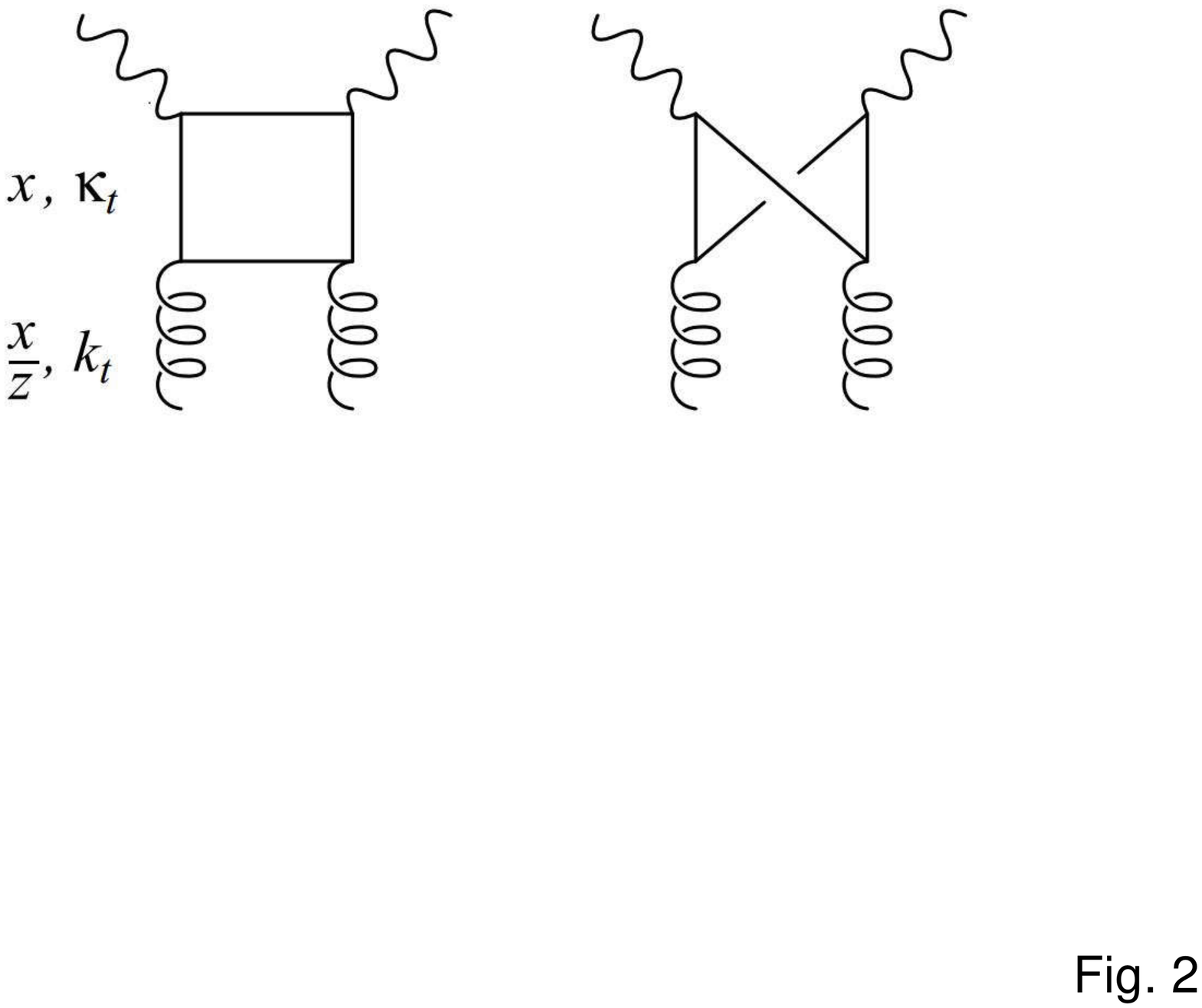}
\caption{The quark and crossed box diagrams, which mediate the
contribution of the unintegrated gluon distribution, $f_g$($x$/$z$,
$k_t^2$, $\mu^2$), to the structure function, $F_2$.}
\end{figure}
\begin{figure}[h!]
  \includegraphics [ scale=0.75]{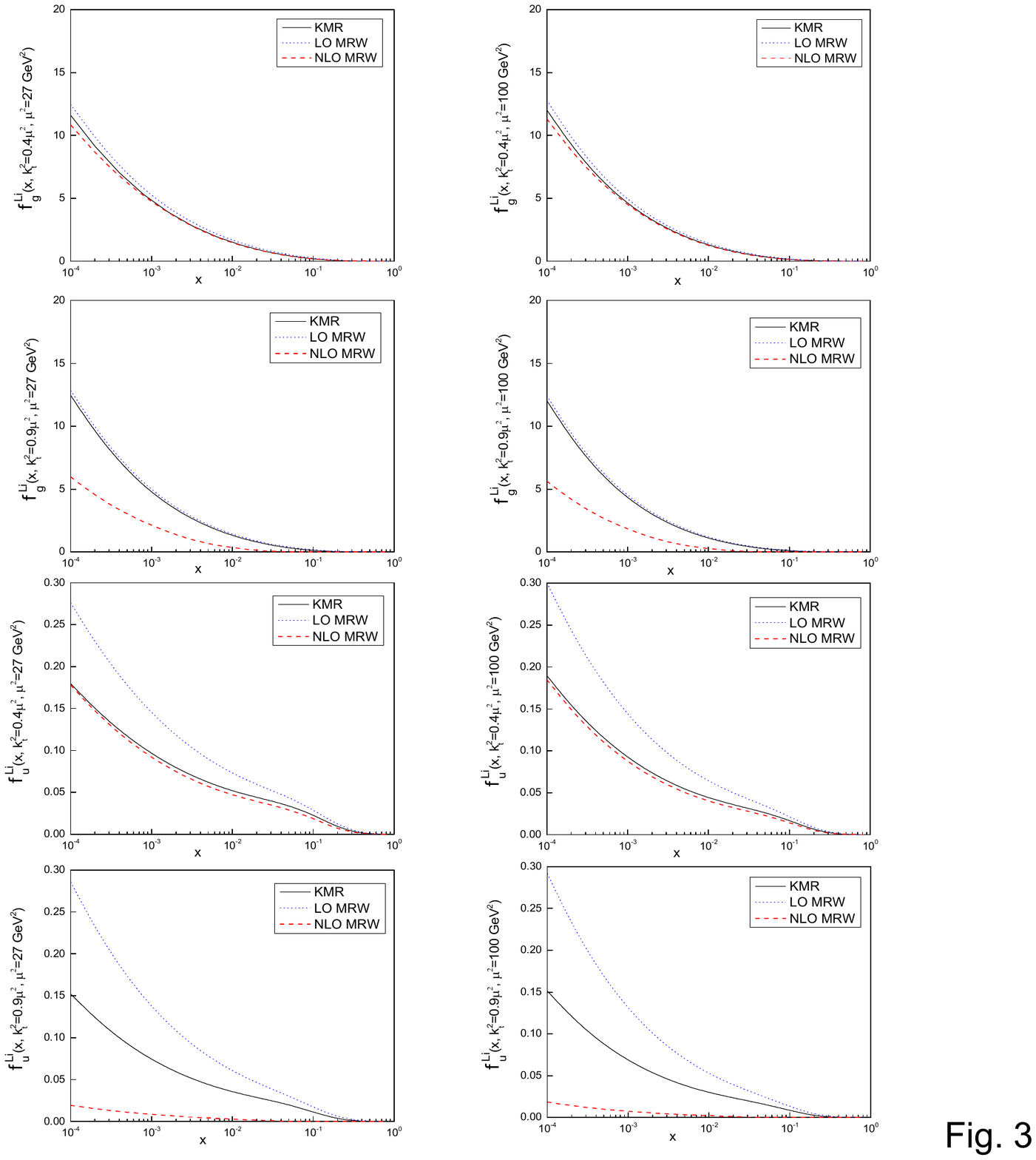}
\caption{The unintegrated gluon and up quark distribution functions
of $^6Li$ nucleus versus $x$, by using the KMR and LO  and NLO MRW
approaches, at the factorization scales $\mu^2$ = 27 $GeV^2$ (the
left panels) and 100 $GeV^2$ (the right panels).}
\end{figure}
\begin{figure}[h!]
  \includegraphics [width=\linewidth]{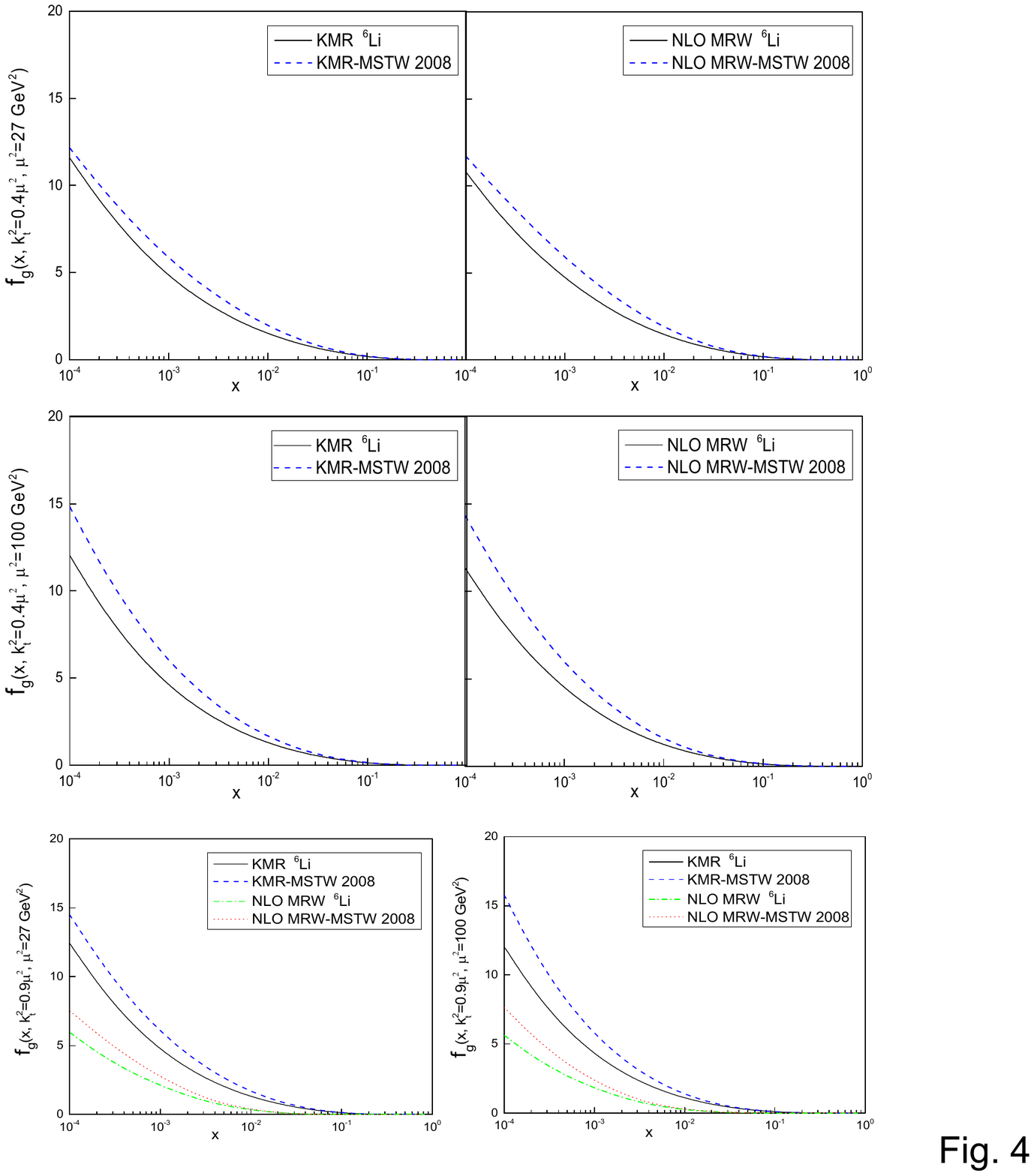}
\caption{The unintegrated gluon distribution functions of $^6Li$
nucleus (present work) and those of the free proton, \cite{Olanj1},
versus $x$, in the KMR and NLO MRW prescriptions, at the
factorization scales $\mu^2$ = 27 and 100 $GeV^2$.}
\end{figure}
\begin{figure}[h!]
  \includegraphics [width=\linewidth]{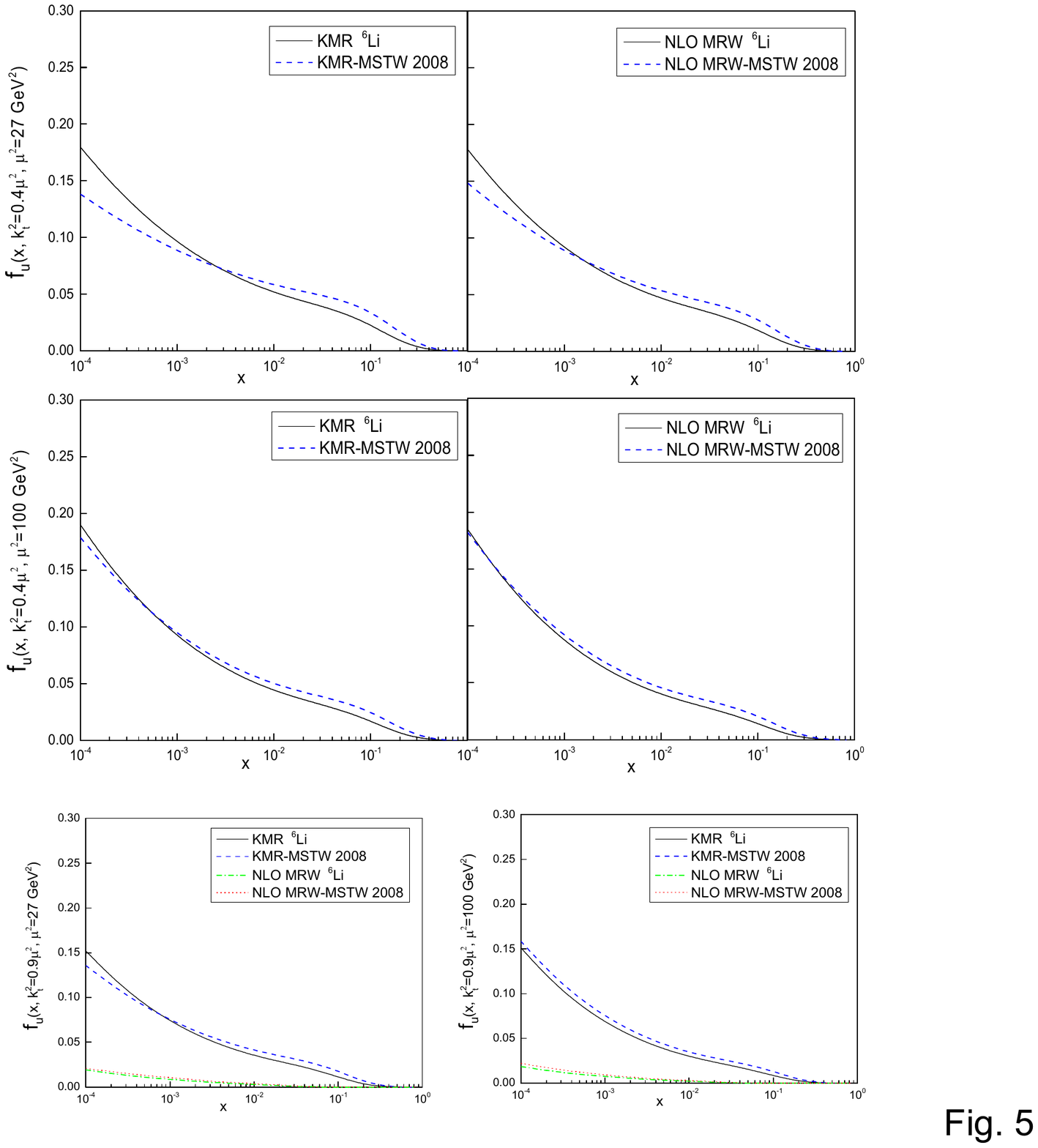}
\caption{The same as the figure 4, but for the unintegrated up
quarks.}
\end{figure}
\begin{figure}[h!]
  \includegraphics [ scale=0.6]{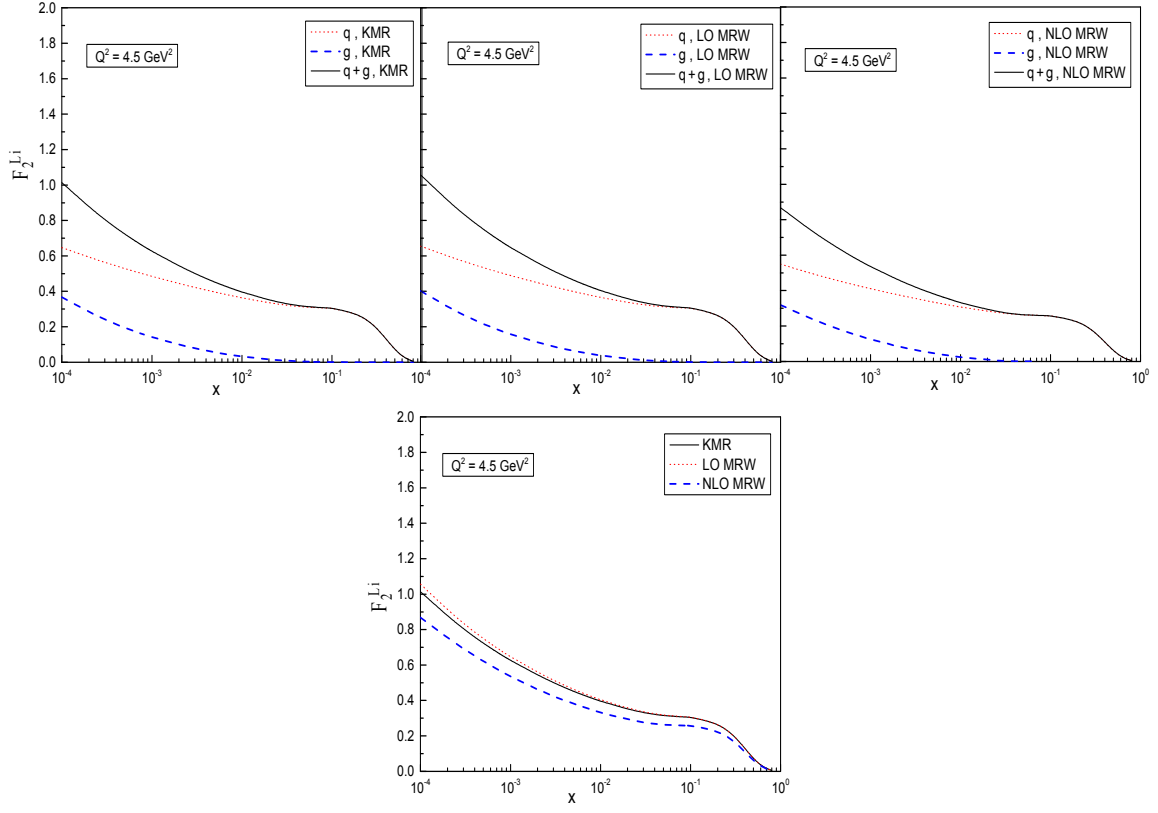}
\caption{The structure functions of $^6Li$ nucleus versus $x$ in the
$k_t$-factorization framework by using the KMR, LO MRW, and NLO MRW
UPDFs, at the factorization scale $Q^2$ = 4.5 $GeV^2$. }
\end{figure}
\begin{figure}[h!]
  \includegraphics [ scale=0.6]{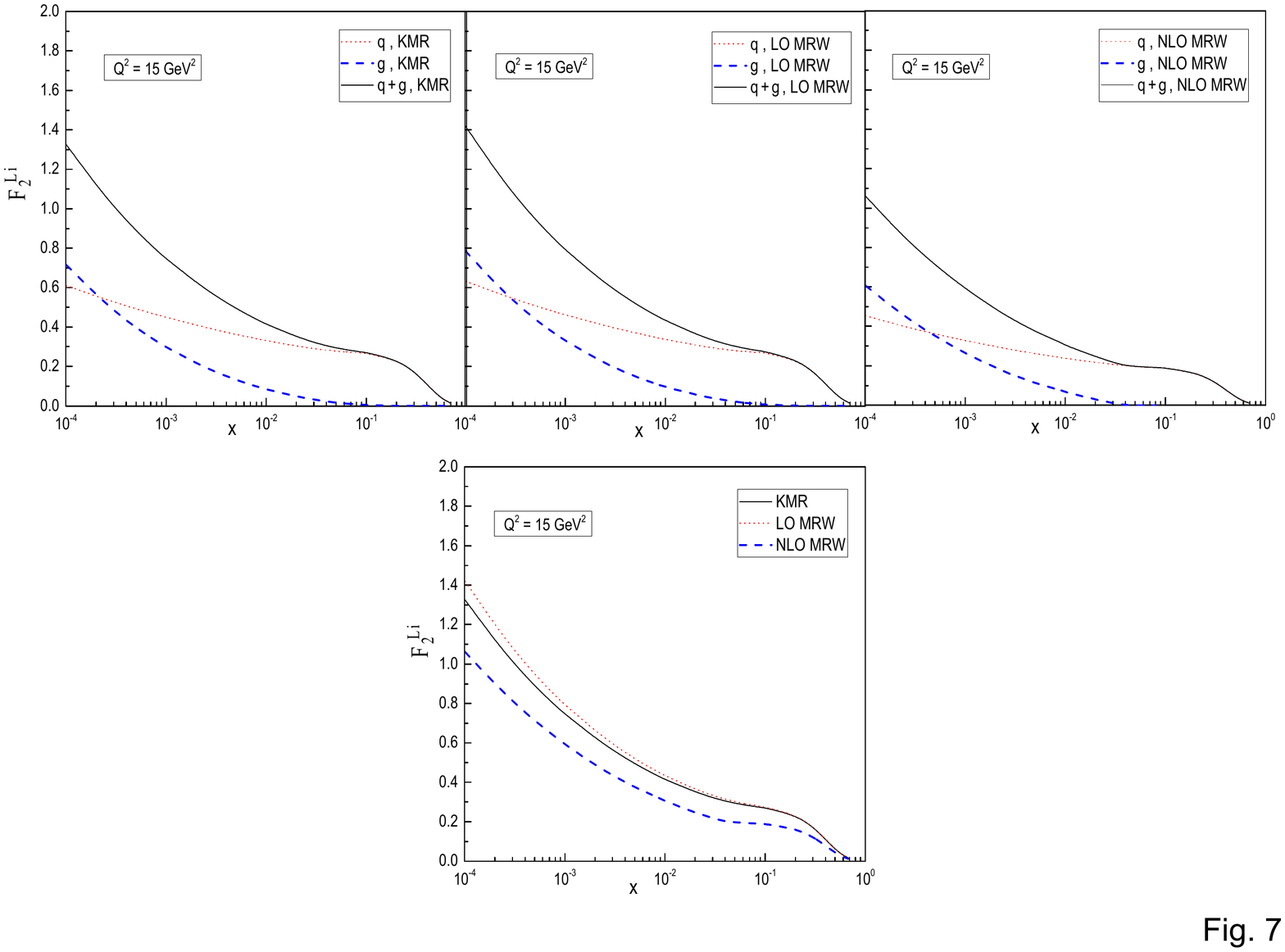}
\caption{The same as the figure 5, but at the factorization scale
$Q^2$ = 15 $GeV^2$. }
\end{figure}
\begin{figure}[h!]
  \includegraphics [ scale=0.6]{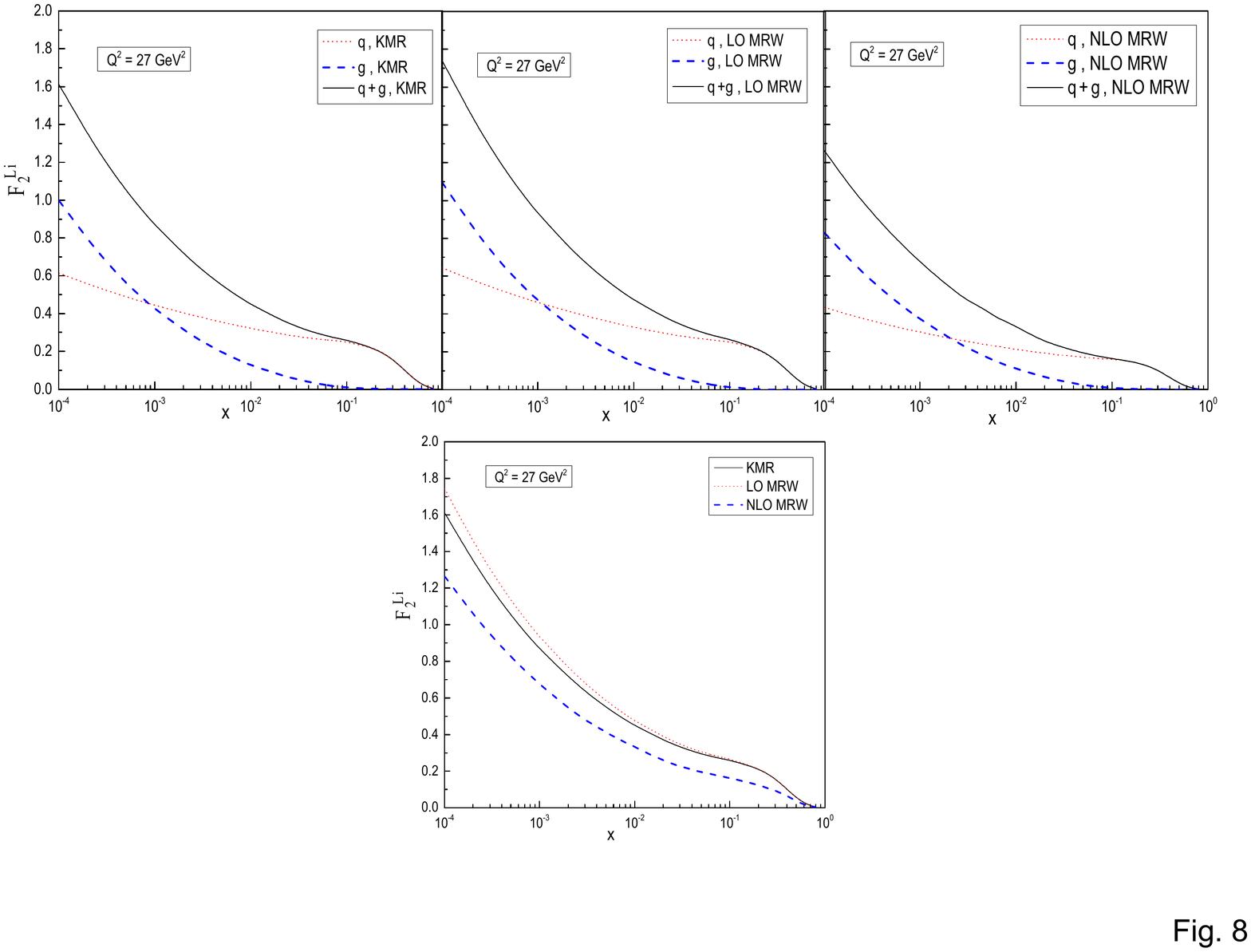}
\caption{The same as the figure 5, but at the
factorization scale $Q^2$ = 27 $GeV^2$.}
\end{figure}
\begin{figure}[h!]
  \includegraphics [ scale=0.6]{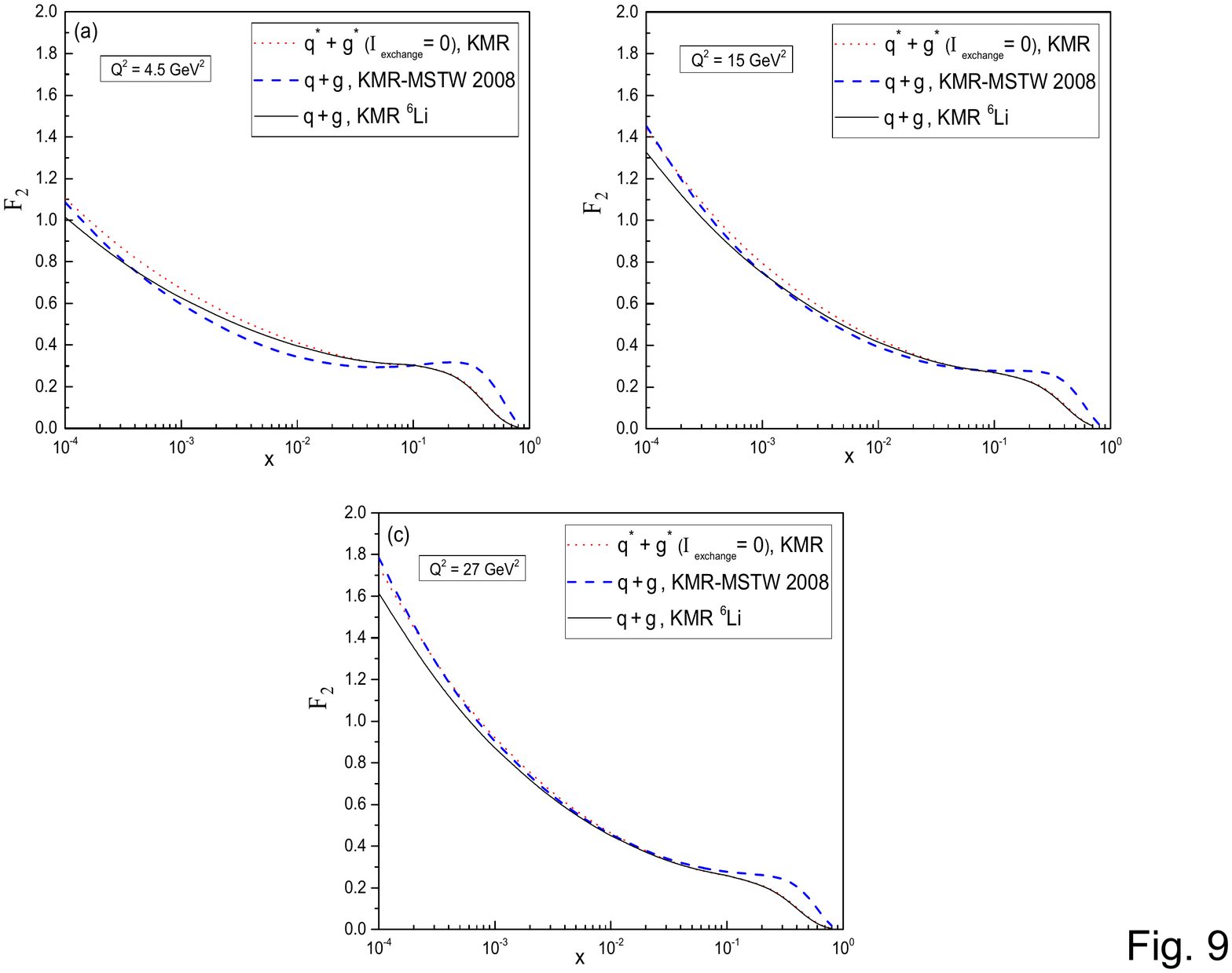}
\caption{The comparison of the SFs of the $^6Li$ nucleus in the KMR
scheme (the full curves) with those of the free proton using the
MSTW 2008 data set as input (the dash curves) at the energy scales
4.5 $GeV^2$ (panel (a)), 15 $GeV^2$ (panel (b)) and 27 $GeV^2$
(panel (c)). The dotted curves represent our hypothetical free
nucleon (by setting the exchange integral $\mathcal{I}$ equal to
zero in the momentum density  of $^6Li$ nucleus, the equation (2))
SFs in the KMR scheme. All SFs are calculated with considering the
three lightest quark flavors ($u$, $d$, $s$).}
\end{figure}
\begin{figure}[h!]
  \includegraphics [ scale=0.6]{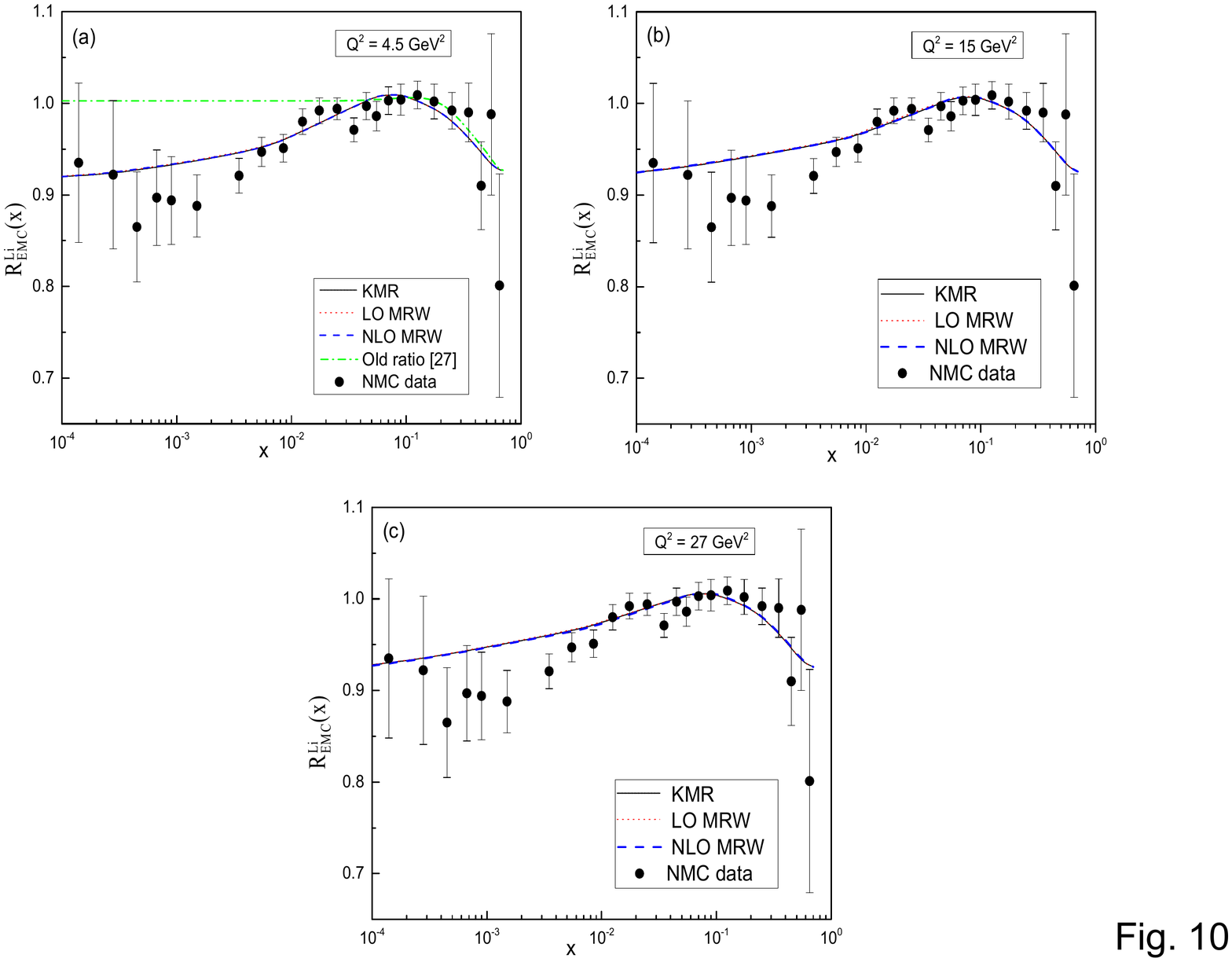}
\caption{The EMC ratio of $^6Li$ nucleus in the KMR (the full
curve), LO MRW (the dotted curve), and NLO MRW (the dash curve)
prescriptions at the energy scales $Q^2$ = 4.5, 15 and 27 $GeV^2$
(note that in the each panel, these curves, especially at larger $x$
values, completely overlap). the circles are the NMC experimental
data \cite{Malace,Arneodo}, and the dotted-dash curve, in the panel
(a), is given from the reference \cite{Hadian1} at $b$ = 0.8 $fm$
and $Q^2$ = 0.34 $GeV^2$, in which the contributions of UPDFs are
not accounted in the EMC calculations.}
\end{figure}
\end{document}